# Evidence for nonlocality and nontemporality of a single photon[*]

A. Driessen

*MESA+ Research Institute, University of Twente, P.O. Box 217, 7500 AE Enschede, The Netherlands*

**Abstract**
*An analysis of the energy exchange by photons is presented based on single-photon Gedanken experiments and the Heisenberg uncertainty principle. Excluding hidden variable properties of a single photon one has to accept that the total photon trajectory undergoes causal influences from the final state separated in time and space from the initial state: nonlocality and nontemporality.*

---

[*] This is the second of a collection of four studies dealing with the weird properties of a photon, compiled in 2017 by Alfred Driessen. The original version of 2003 has not yet been published and the current version is essentially unchanged. Only incidentally an update of the references is given.

The collection includes:

    1. Fundamental lower size limit in wavelength selecting structures,
A. Driessen, H.J.W.M. Hoekstra, D.J.W. Klunder and F.S. Tan, 2003, unpublished
published in 2017: https://arxiv.org/ftp/arxiv/papers/1707/1707.05180.pdf

    2. Evidence for nonlocality and nontemporality of a single photon,
A. Driessen, 2003 (unpublished),
published 2017 in arXive

    3. Propagation of short lightpulses in microring resonators: Ballistic transport versus interference in the frequency domain,
A. Driessen, D.H. Geuzebroek, E.J. Klein, R. Dekker, R. Stoffer, C. Bornholdt
Optics Communications 270 (2007) 217-224. doi:10.1016/j.optcom.2006.09.034
https://www.researchgate.net/publication/221661942_Propagation_of_short_lightpulses_in_microring_resonators_Ballistic_transport_versus_interference_in_the_frequency_domain

    4. The strange properties of the photon: a case study with philosophical implications,
Alfred Driessen, Talk presented at Journée « Science, raison et foi » on Causalite, Temps et Origine de l'Univers, CASTELVIEIL, MARSEILLE, November 12 , 2013,
https://www.slideshare.net/ADriessen/the-strange-photon

**1. Introduction**
The most puzzling phenomenon in physics is light, or emphasizing its particle character, the photon. This stems from the fact that it is the particle with the strongest quantum character and simultaneously the most relativistic one. Both, quantum mechanics and the theory of relativity are frameworks that seem to contradict nearly completely our every-day-life experience[1],[2]. As an example one may mention the EPR paradox[3] and the introduction of quantum nonlocality in the subsequent scientific discussion[4].

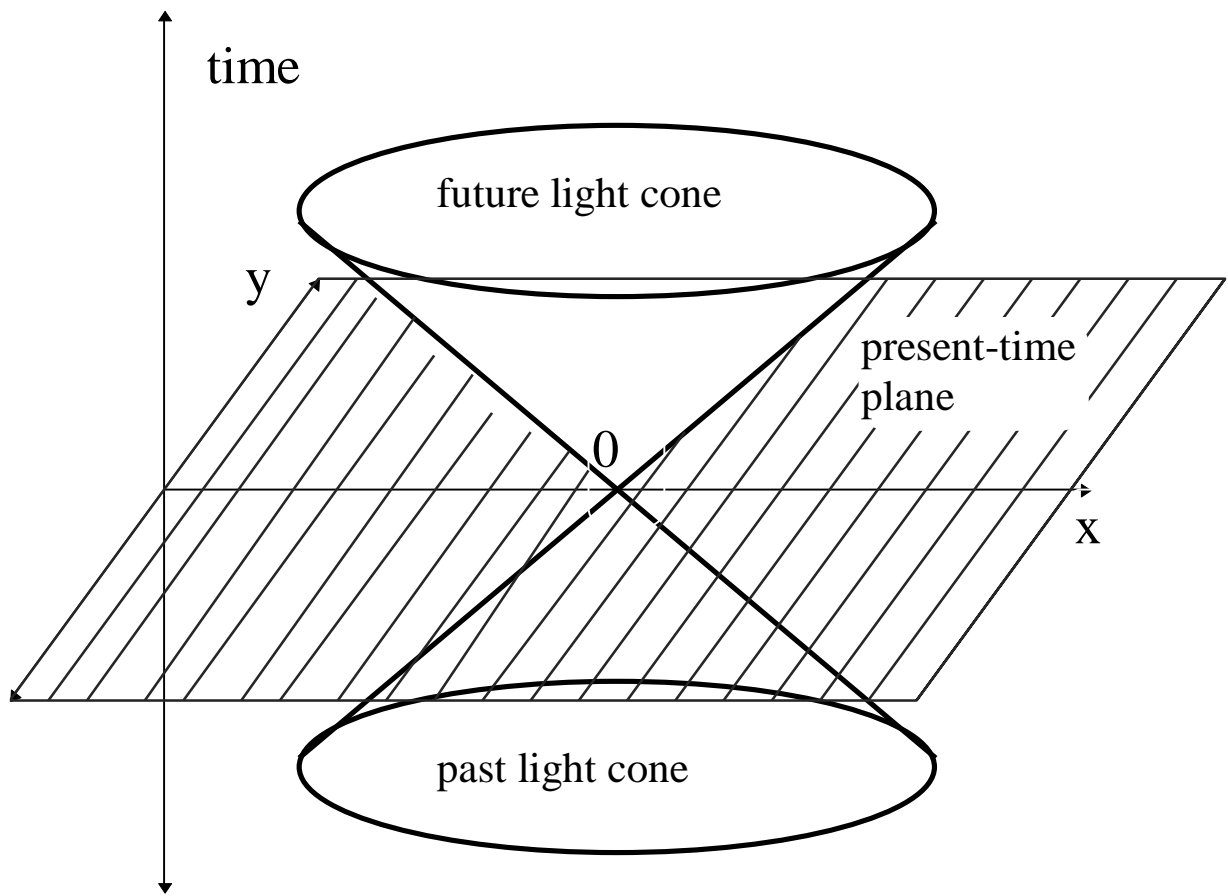

Figure 1: Space-time diagram (2 space coordinates *x-y*; and the time coordinate *t* ) for an event located at the origin.

Nonlocality and nontemporality refer to phenomena where one encounters influences of causes that are separated in space and time in a peculiar way. Fig. 1 shows an *x-y-t* diagram (two space- and one time-coordinate) with the present-time plane and the future and past light cones of an event located at the origin. In classical physics and relativity causes of the event lie in the past light-cone, they are connected by a time-like interval with the event. Nonlocality means that causes are located in the past half space or in the generalized present but outside the past light cone; the interval connecting cause and event is spacelike. As a consequence faster than light effects are involved or action on a distance. Nontemporality means that events in the future light cone have a certain influence on the present state on the origin; the interval connecting cause and event is timelike but with the arrow of time reversed. In the case of nontemporality faster than light effects would not solve the striking causal relation.

The basic idea of a photon can be summarized as follows: a photon is a transfer of a quantum of energy *E* from one place *A* to another place *B*. The speed of energy transfer is finite and has in vacuum for all energies a constant value *c*, which is called the speed of light. The positions of all possible end places *B* at a given time define a surface around the source *A* expanding with the speed of light. The volume enclosed by this surface is related to the concept of the so-called "wavefunction" that "collapses" eventually at a single site *B*. This volume is determined by the geometry and optical properties of objects and can be highly a-symmetric and bizarre.

The trajectory of a single photon from *A* to *B* is not directly accessible by any experimental method and can only be approached by statistical methods. In the case of large numbers of photons, approximate models like the ray-picture or Maxwell Equations solved by, e.g. Beam propagation algorithms, give good results[5]. In the individual case one can work only with wavefunctions, whose amplitude squared is proportional to the probability to detect a photon at that peculiar position. Single photon experiments show that photons can follow in a certain sense simultaneously parallel paths and have an extension, the coherence length. They show particle properties that are definitely different from those of macroscopic objects.

In the following argumentation about nontemporality often the Heisenberg uncertainty principle (HUP) will be applied to single photons. This principle can be stated as[6]:

$$\Delta E \, \Delta t > h/(2\pi) \tag{1}$$

or in terms of the frequency ν of the photon:

$$\Delta \nu \, \Delta t > 1/(4\pi) \tag{2}$$

where Δ*E* and Δ*t* are the uncertainty in energy and time respectively of the photon and *h* the Planck constant; alternatively one can write:

$$\Delta \mathbf{x} \Delta \mathbf{p} > \hbar/2 \tag{3}$$

where *x* is the place and *p* the momentum of the photon. In the case of single photons in a Fourier transform limited pulse the ">" symbol in Eqs. (1), (2) and (3) can be changed to "=".

In the HUP in the formulation of Eq. (1) the uncertainty in time $\Delta t$, or the uncertainty in length, the coherence length $l_c = c\Delta t$ is in general referred to as the pulse-duration and the coherence length, respectively, of the source[7]. The aim of this study is to present evidence that $\Delta t$ or $l_c$ of a photon firstly is determined by any principal time uncertainty in the whole trajectory including source, single, multiple or parallel paths and the detector arrangement and, secondly, is unchanged during the total trajectory from source *A* to endpoint *B*.

**2. Nontemporal processes involved in the interaction by photons**
In the following it will be shown that nontemporal processes play an important role in the energy exchange by photons; for recent related work, see, e.g.,[8]. The argumentation is based on three main ideas: (i) the coherence length of a photon remains unchanged during the complete lifetime from emission to absorption; (ii) the coherence length is dependent not only on the source properties but on the whole trajectory (iii) the trajectory of a photon is dependent on the coherence length. The consequences are dramatic, as photons of astronomical light sources connect events separated in space and time by billions of light-years, and billions of years respectively. For the support of the argumentation, Gedanken experiments are performed that exclude some straightforward but probably too naïve conclusions.

Before starting the discussion one has to clarify whether the photon is an object of the physical reality. In the case of a full no, one is confronted with the experimental fact that at one place *A* and time $t_i$ energy vanishes and at a certain time later at place *B* and time $t_f$ energy is generated. If there is no physical reality connecting those two events in space and time, then one is forced to accept action-at-distance in space as well as in time. Also the law of conservation of energy would not be obeyed during the time between emission and absorption. At astronomical distances between places *A* and *B* the timescale of missing energy could be as large as millions of years. One therefore is compelled to accept that the photon is an object of the physical reality. It is relevant to note that also in the classical view, as expressed in Maxwell's equation, there is a physical reality connecting time and place of emission and absorption of electromagnetic radiation, namely the electromagnetic field.

The argumentation leading to evidence for the nontemporal character of a single photon is not trivial and requires a series of argumentations with side-branches and direct as well as indirect proofs, see the flow-chart of Fig. 2. Central in the discussion are a series of Gedanken-experiments with single photons. In the case of linear optics this is not a real restriction as in that case, by definition, there is no interaction between photons. In order to arrive at meaningful results all experiments are repeated sufficiently so that statistical noise can be neglected. The photon source could be the spontaneous emission of a dilute cold gas producing photons with energy $E = h\nu_0$. The spread in frequency $\Delta\nu$ is only determined by the natural line width and is related to the lifetime $\tau_R$ of the excited state by the HUP (2) which reads in this case:

$$\Delta\nu\,\tau_R = 1/(4\pi) \qquad (4)$$

The analysis of photons can be done by the following ideal apparatuses:
1) the ideal high speed photo detector with 100% quantum efficiency at all wavelengths of interest and small detection area;
2) the ideal, loss-less spectrometer, based on, e.g. a prism with *N* output channels;
3) the ideal, loss-less Michelson interferometer with one of the mirrors placed on a translation stage for the determination of the coherence length.
4) the ideal loss-less single mode fiber with non-zero dispersion

With this equipment we start our first experimental analysis of the photon source. The spectrometer together with a detector-array allows the determination of *ν* and Δν of the photon source. In addition, by measuring the output intensity of the Michelson interferometer as a function of the translation of the interferometer mirror, the coherence length or coherence time $t_c = l_c/c$ of the photon source can be determined. The coherence time measured is related to a characteristic time scale of the photons arriving at the detector. In this specific experiment the only relevant time-scale for the photon is the lifetime of the photon at the source $\tau_R$. The experiment gives therefore: $t_c \approx \tau_R$. One finds that Eq. (2) is fulfilled.

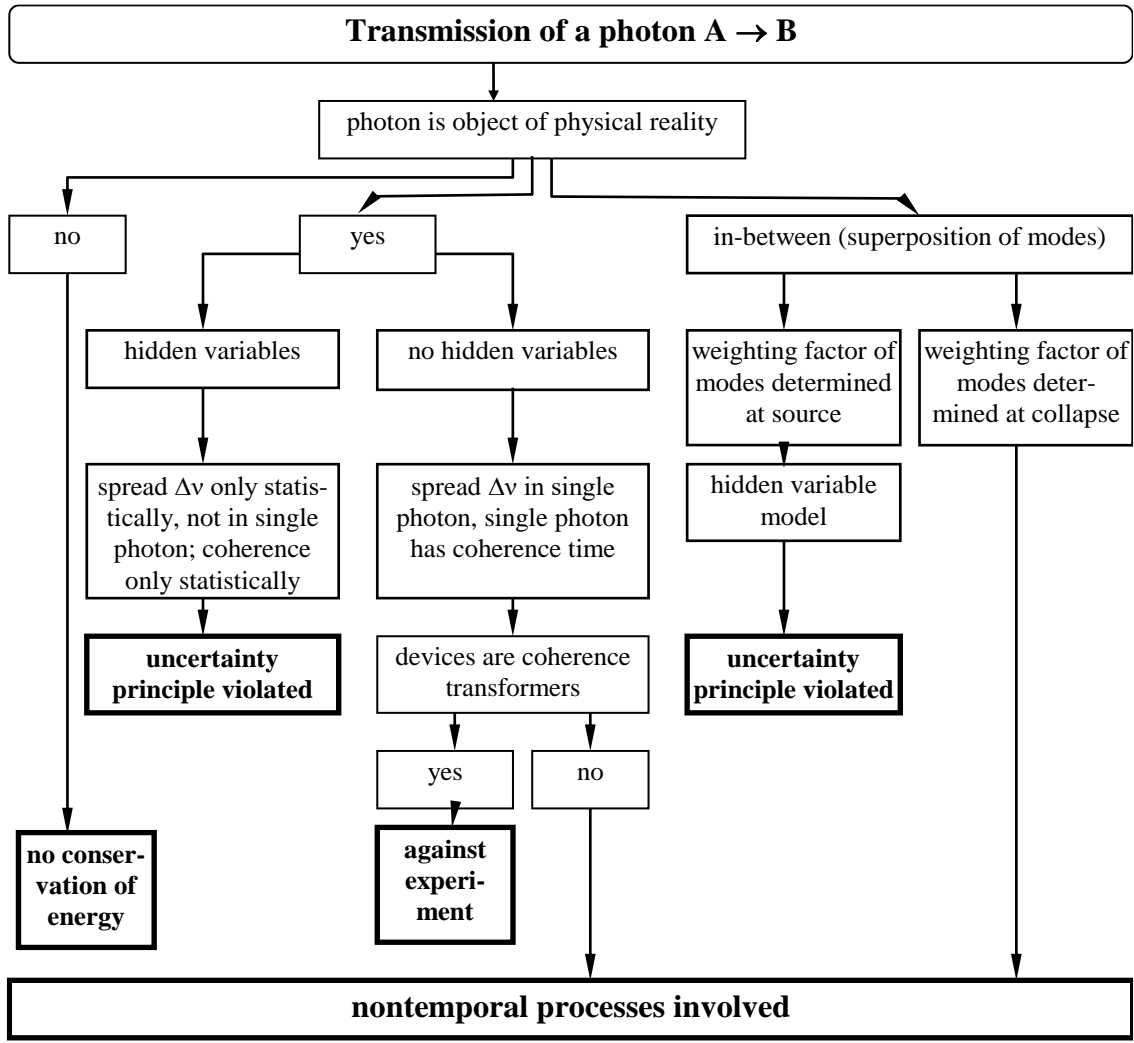

Fig. 2: 'Flow-chart' of argumentation used in section 2

In the next Gedanken experiment photons of the above-described source are passed through a spectrometer in order to separate in space photons with different frequency. Each photon is directed to one of the $N$ output channels. Connecting each channel with a photo detector the frequency $v$ and spread in frequency $\Delta v$ can be measured with the same result as in the first experiment. When the Michelson interferometer is used to measure the coherence length of the photons at any of the output channels of the spectrometer, one finds a much larger coherence length $l'_c$ than in the first experiment. Putting an additional spectrometer at any of the output ports of the first spectrometer one can measure a much reduced spread in frequency $\Delta v'$ and the Heisenberg uncertainty relation, Eq. (2), now becomes (with $t'_c$ the new coherence time):

$$\Delta v' \, t'_c = 1/(4\pi) \tag{5}$$

The characteristic time-scale of the photon has changed, as in addition to the lifetime of the source an additional time-uncertainty occurs in any spectrometer due the uncertainty in the geometrical path length of the photon[9].

One therefore can conclude that the same source can provide - depending on the experiment - groups of photons with different intrinsic properties, namely a different coherence length and different spread in frequency. The question now is, whether we really deal with different properties. Is it not possible that in the second experiment the photons are selected according to their frequency? Such a subgroup would automatically have a larger coherence length and reduced $\Delta v$. In the first experiment we could not detect the larger coherence length as, due to the spread in frequency, the interference pattern at the output Michelson interferometer has been smeared out. If this assumption would be true, then our photon source would emit individual monochromatic photons. The spread in frequency would only appear in a group of photons. In this picture we could not observe the longer $t_c$' as each photon had its individual longer coherence length which in our measurement could not be observed. The reason for this was our statistical method, which always treated photons with a rather large spread in frequency $\Delta v$.

This assumption can be reduced to a hidden variable theory, as it ascribes precise, but not measurable, values to the single photon. As a consequence one has to give up the HUP for the individual photon. If the photon source would emit photons with each photon having a very low spread in frequency and consequently a large coherence time, then Eq. (2) is not fulfilled. Accepting quantum mechanics as a valid theory, the hidden variable approach has to be rejected.

There are two other possibilities left for explaining the fact that two different values for the coherence length can be obtained with photons from the same source. The first is that a spectrometer acts as a coherence length transformer. The alternative is that the properties of the photon emanating from the photon source depend on the circumstances the photon will encounter on its trajectory. In order to get deeper insight in these alternatives we perform a third Gedanken experiment. The light of our light source now is directed to a single mode fiber with a certain finite dispersion. The length of the fiber is chosen long enough so that the dispersion induced spread in transmission time $\Delta t_{fiber}$ exceeds by far the lifetime of the source, i.e. $\Delta t_{fiber} >> \tau_R$.

After transmission through the fiber the photons are directed to a spectrometer, whose $N$ output channels are each connected with a photo detector. The same measurements as in the second experiment are now performed. In addition, the time $t_N$ between excitation of the source and detection at the photo detector for all channels $N$ is measured. Within the coherence length transformer picture one would expect nearly the same $<t_N>$ for all $N$. In the other case one would see a spread in $<t_N>$ nearly equal to $\Delta t_{fiber}(v)$. Experimentally one finds that the photons of each channel $N$ have their own $t_{fiber}(v)$ in accordance with the fiber dispersion for a photon with frequency $v$ and the reduced spread in $\Delta v$. As a conclusion, there is no coherence length transformation. The photons entering the fiber have already a coherence length different from the photons of the bare source that is not connected to a spectrometer. This is in accordance with an everyday experience in linear optics: an optical filter (like the spectrometer) can be placed directly behind the source or just in front of the detector without changing the result.

Are we not contradicting our own argumentation given above by stating that the spectrometer is just distributing in space the photons that already had their individual long coherence length? We rejected that argument by claiming that the HUP should be fulfilled. The only explication could be that the HUP Eq. (2) does refer not only to principal uncertainties in time at the source but also at any part of the whole trajectory. The trajectory has to be considered as a whole. The photon has a unique and unchanged coherence length and uncertainty in energy all over the whole trajectory from source to absorber. The values of these, however, are not determined by the source alone, but also by any part of the trajectory that introduces a principal time uncertainty. In the third Gedanken experiment, for example, the coherence length determining the duration of the photon transmission through the fiber was also dependent of the apparatus - the spectrometer - placed nearly at the end of the trajectory. The photons probe the presence of the spectrometer only in the last phase of the trajectory, the causal influence, however, is already present in the first phase of the trajectory. Therefore, with the observation of the Gedanken experiment that the photon trajectory depends on the basic photon properties, i.e. the coherence length and the spread in energy, one arrives at the conclusion that the transmission of a photon is connected to a nontemporal process.

Some brief remarks should be made for the case that the photon is not considered as a clear and distinguished object of the physical reality but something having only a strong relation with reality. This is the widely used approach of the superposition of modes[10]. After creation of the photon at *A*, the possible trajectories from *A* to *B* are described in photon modes that develop according to Maxwell's equations. These modes together constitute the wavefunction, that at a certain moment collapses at site *B*. The expansion in modes is a common technique but the question remains, what is determining the weighting factor of each mode. If it is determined at the source, then we have a hidden variable theory where the arguments given above are valid. If the weighting factors are determined at the moment of the collapse, then the weighting factors of all not relevant modes have to be set zero already at the emission event. Finally if energy packets propagating in the different modes are considered only as mathematical objects then we are back to the "no physical reality" option that we excluded in the beginning of this section. Therefore also the model of superposition of modes leads eventually to the acceptance of nontemporal processes.

After having gone through the branches of the argumentations outlined in Fig. 2 the conclusion becomes obvious: in the case of energy-exchange by photons nontemporal processes play a decisive role. In the special case that one considers photons as objects of reality evidence is given that the coherence length is unchanged along the photon trajectory and is determined by the fundamental time uncertainty in the trajectory as a whole.

## 3. Discussion
The conclusion about nontemporality in quantum processes is not new; Sommerfeld already stated in 1930[11]: *When I sometimes spoke about a new and conditioned causality, it was based mathematically on the fact that we had to calculate the radiation of an atom on the basis of a formula where the initial and final states were included equally and symmetrically. That means that in the case of radiation a foresight on the final state together with a memory of the initial state is present as a mathematical fact.* A more recent application of this approach and experimental verification can be found in Snoeks et al.[12] who analyze a well-known phenomenon in waveguide optics in terms of Fermi's golden rule. What is new in the present contribution is the argumentation demonstrating that the influence of the final state applies also to timelike intervals that can be as large as years in the case of astronomical light sources.[†]

In a recent experimental paper on quantum correlations with photons, Stefanov et al.[8,13] conclude that the observed quantum correlations *not only are independent of the distance, but also it seems impossible to cast them in any real time ordering*. This is agreement with the comment by Sommerfeld and our observation in the fiber Gedanken experiment that a change at the end of the photon trajectory has an influence also on the part already completed. It is not difficult to work out delayed-choice Gedanken experiments where the detector arrangement is changed after emission of the photon. Even in that case the new arrangement will not work as a "coherence length transformer", as the above given argumentation of the "static" Gedanken experiments is still valid in the case of delayed choices.

The above given evidence and comparison with related experiments lacks still a direct experimental proof of the nontemporal behavior of single photons. It is highly desirable to conceive an experiment with photons where the outcome is depending on the final state, i.e. depending how the photons are detected. Recent experiments[14] with fs pulses in microring resonators[15] and time resolved detection seem to indicate that in fact there is a difference in the response of the resonator depending on a principle time uncertainty in the detector[16].

In conclusion, evidence is given that nontemporal processes are involved in the energy exchange by photons. The evidence is only indirect but based on experimentally supported principles, like Heisenberg's uncertainty principle. The acceptance of nontemporality is to our knowledge not against any experimental fact. Therefore it is to prefer to models like hidden variable theories that are likely to be in conflict to experiment[17].

---

[†]Note added in 2017:
The special relation of a photon with time is emphasized by Roger Penrose in a study based on relativity and quantum mechanics:
--*The point is that, according to a massless particle, the passage of time is as nothing.* [R. Penrose, *Cycles of Time,* Random House, London 2010, p. 146].
--*Eternity is no time at all, for a photon.* [R. Penrose, slide presented during a public talk at University of Leiden, 10-6-2011].

In the introduction has already been stated that nontemporality refers to phenomena where one encounters influences of causes that are separated in time in a peculiar way. If there is a reference frame for the photon where there is no time at all between the point of emission and absorption then a new situation arises. In that system the final state is not in the future but at present with the initial state and any other state of the trajectory. Penrose's observation is therefore compatible with the Gedanken experiments described in section 2.